\documentclass[aps,superscriptaddress,showpacs,nofootinbib,eqsecnum]{revtex4}

\usepackage{amssymb}
\usepackage{amsmath}
\usepackage{natbib}
\usepackage{epsfig}
\usepackage{graphicx}
\usepackage{dcolumn}
\usepackage{bm}
\usepackage[english]{babel}
\usepackage[latin1]{inputenc}

\begin{document}

\title{Applicability of the Linear $\delta$ Expansion for the $\lambda \phi^4$
  Field Theory at Finite Temperature in the Symmetric and Broken Phases}

\author{R. L. S.  Farias} \email{ricardo@dft.if.uerj.br}
\affiliation{Departamento de F\'{\i}sica Te\'orica, Universidade do Estado do
  Rio de Janeiro, 20550-013 Rio de Janeiro, RJ, Brazil}

\author{G. Krein} \email{gkrein@ift.unesp.br} \affiliation{Instituto de
  F\'{\i}sica Te\'orica, Universidade Estadual Paulista, Rua Pamplona 145,
  01405-900 S\~ao Paulo, SP, Brazil}

\author{Rudnei O.  Ramos} \email{rudnei@uerj.br} \affiliation{Departamento de
  F\'{\i}sica Te\'orica, Universidade do Estado do Rio de Janeiro, 20550-013
  Rio de Janeiro, RJ, Brazil}

\begin{abstract}
  
  The thermodynamics of a scalar field with a quartic interaction is studied
  within the linear $\delta$ expansion (LDE) method. Using the imaginary-time
  formalism the free energy is evaluated up to second order in the LDE. The
  method generates nonperturbative results that are then used to obtain
  thermodynamic quantities like the pressure. The phase transition pattern of
  the model is fully studied, from the broken to the symmetry restored phase.
  The results are compared with those obtained with other nonperturbative
  methods and also with ordinary perturbation theory.  The results coming from
  the two main optimization procedures used in conjunction with the LDE
  method, the Principle of Minimal Sensitivity (PMS) and the Fastest Apparent
  Convergence (FAC) are also compared with each other and studied in which
  cases they are applicable or not. The optimization procedures are applied
  directly to the free energy.

\end{abstract}

 \maketitle
\section{Introduction}

Phase transition phenomena in quantum field theories are typically of
nonperturbative nature and thus naive perturbation theory based on an
expansion in the coupling constant cannot be employed. This is clearly the
case of phase changes at high temperatures, where perturbation theory becomes
unreliable because powers of the coupling constant become surmounted by powers
of the temperature~\cite{lebellac}.  Problems with perturbation theory also
happen in phenomena occurring close to critical points, because large
fluctuations can emerge in the system due to infrared divergences, thus
requiring nonperturbative methods as well in their studies.  This is the case
of studies involving second order phase transitions and also in weak first
order phase transitions~\cite{GR}. Typical examples where these problems can
manifest are in studies of symmetry changing phenomena in a hot and dense
medium, a subject of interest in quantum chromodynamics (QCD) in the context
of heavy-ion collision experiments, and also in studies of the early universe.
Consequently, there is a great deal of interest in investigating thermal field
theories describing matter under extreme
conditions~\cite{revQGP,rev_dirkie,revrebhan,revanders_strick}.

{}Familiar nonperturbative methods that have been used to study symmetry
changing phenomena at finite temperatures are resummationlike techniques,
such as the daisy and superdaisy schemes~\cite{Espinosa,sato}, composite
operator methods~\cite{Camelia}, and field propagator dressing
methods~\cite{Banerjee,Parwani}. Other methods used include expansions in
parameters not related to a coupling constant, like the $1/N$ expansion and
the $\epsilon$-expansion~\cite{zinn}.  In addition, there are numerical
methods, the most notably ones are those based on lattice Monte Carlo
simulations~\cite{lattice}. Each method has its own advantages and
disadvantages. {}For instance, in numerical methods there may be issues
related to numerical precision, lattice spacing, and lattice sizes.
In~addition, there is the notorious problem of simulating fermions on the
lattice at finite chemical potentials~\cite{lattice}. In any nonperturbative
method based on an expansion in some parameter one has to face the problem of
higher order terms becoming increasingly cumbersome, so stalling further
analysis. This is usually the case when carrying out calculations beyond
leading order in the $1/N$ expansion. Careless use of a nonperturbative method
can also lead to problems like the lack of self-consistency or overcounting of
effects. Known examples of such problems are the earlier resummation works
dealing with daisy and superdaisy schemes, that at some point were giving
wrong results, e.g. predicting a first order transition~\cite{carr} for the
$\lambda \phi^4$ theory, an unexpected result since the model belongs to the
universality class of the Ising model, which is second order.  These methods
also predicted a strong first order phase transition in the electroweak
standard model, a result proved to be misleading~\cite{arnold}.

Let us recall that the breakdown of perturbation theory at high temperatures
and its poor convergence properties have been dealt with many different
methods.  Examples are the use of self-consistent
approximations~\cite{gruter+alkofer}, hard-thermal-loop (HTL)
resummation~\cite{htl,htl12345}, perturbative expansions in the coupling
constant with resummation implemented with the use of a variational mass
parameter, also known as screened perturbation theory
(SPT)~\cite{spt,andersen-braaten-strickland}, and the use of two-particle
irreducible (2PI) effective actions~\cite{cjt}. The 2PI method, in particular,
leads to a much better convergence of thermodynamic quantities (like the
pressure) as compared to some of the other methods~\cite{berges+reinosa}.
Related to the 2PI method is the $\Phi$-derivable technique, which has been
used to study the thermodynamics of scalar and gauge
theories~\cite{iancu1,iancu2,iancu3,peshier}.  One difficulty with the 2PI
effective actions is that the renormalization procedure is
nontrivial~\cite{knoll_urko_b2_fejos}. In addition, there seems that the
$\Phi$-derivable technique breaks down for a coupling beyond some
value~\cite{braaten+Peti}.

In general, it is desirable that any analytical nonperturbative method obey
two basic requirements. First, it should be self-consistent, and second, it
should produce useful results already at lowest orders without the need for
going to higher orders. That is, it should produce results that quickly
converge at some order where calculations are still feasible analytically or
semianalytically. Though some of the cited methods may satisfy one, or to
some extent both of these requirements, in the present paper we are
particularly interested in the one known as the linear $\delta$ expansion
(LDE)~\cite{linear}, a nonperturbative method that has been used successfully
in different contexts related to thermal field
theories~\cite{GN,marcus-rudnei,mra} and in many other theories -- for a long,
but far from complete list of references see Refs.~\cite{early,KMP}. In the
LDE, a linear interpolation on the original model Lagrangian density is
performed in terms of a fictitious expansion parameter $\delta$, which is used
only for bookkeeping purposes and set at the end equal to one.  The standard
application of the LDE to a theory described by a Lagrangian density
$\mathcal{L}$ starts with an interpolation defined by
\begin{eqnarray}
{\cal L} \to \mathcal{L}^{\delta } &\mathcal{=}&
\left( 1-\delta \right) \mathcal{L}
_{0}\left( \eta \right) +\delta \mathcal{L} \nonumber\\
&=&\mathcal{L}_{0}\left( \eta \right) +\delta \left[ \mathcal{L-L}_{0}\left(
\eta \right) \right]\;,
\label{opt}
\end{eqnarray}
where $\mathcal{L}_0$ is the Lagrangian density of a solvable theory, which is
modified by the introduction of an arbitrary mass parameter (or parameters)
$\eta$. The Lagrangian density $\mathcal{L}^{\delta}$ interpolates between the
solvable $\mathcal{L}_0(\eta)$ (when $\delta=0$) and the original
$\mathcal{L}$ (when $\delta = 1$). The procedure defined by Eq. (\ref{opt})
leads to modified {}Feynman vertices, that become multiplied by $\delta$, and
modified propagators, that now depend on $\eta$. All quantities evaluated at
any finite order in the LDE will then depend explicitly on $\eta$, unless one
could perform a calculation to all orders. Up to this stage the results remain
strictly perturbative and very similar to the ones obtained via an ordinary
perturbative calculation. It is through the freedom in fixing $\eta$ that
nonperturbative results can be generated in this method.  Since $\eta$ does
not belong to the original theory, one may fix it requiring that a physical
quantity $\Phi^{(k)}$, calculated perturbatively to order-$\delta^k$, be
evaluated at the value where it is less sensitive to this parameter. This
criterion, known as the principle of minimal sensitivity (PMS), translates
into the variational relation~\cite{PMS}
\begin{equation}
\frac {d \Phi^{(k)}}{d \eta}\Big |_{\bar \eta, \delta=1} = 0 \;.
\label{pms}
\end{equation}
The optimum value $\bar \eta$ which satisfies Eq.~(\ref{pms}) is a function of
the original parameters of the theory. In particular, $\bar \eta$ is a
nontrivial function of the couplings and because of this nonperturbative
results are generated. Another optimization procedure used is known as the
fastest apparent convergence (FAC) criterion~\cite{PMS}. It requires from the
$k$-th coefficient of the perturbative expansion
\begin{equation}
\Phi^{(k)} = \sum_{i=0}^k c_i \delta^i  \;,
\label{fac0}
\end{equation}
that
\begin{equation}
\left[\Phi^{(k)} - \Phi^{(k-1)}\right]\Bigr|_{\delta=1} =0\;,
\label{fac}
\end{equation}
which is just equivalent to taking the $k$-th coefficient (at $\delta=1$) in
Eq.~(\ref{fac0}) equal to zero.

One should note that it is not at all guaranteed that the condition in
Eq.~(\ref{pms}) has a nontrivial solution. In cases where this may happen, the
second criterion, Eq.~(\ref{fac}), may be more appropriate. One example where
the condition given by Eq.~(\ref{pms}) fails to produce a nontrivial solution
was in the problem studied by the authors in Ref.~\cite{senise}, who applied
the LDE to compute the effective potential in superspace.  There, the authors
found that while the PMS condition was unable to give a nonperturbative
solution to the effective potential, the FAC criterion worked perfectly well.
Of course, in many situations both optimization criteria may work and in this
case one may ask whether they lead to equivalent results.  Previous studies
indicated that this is indeed so, but a full comparison of results obtained
with both optimization criteria is still lacking.  Another issue associated
with the LDE is its convergence.  Rigorous LDE convergence proofs have been
obtained for the problem of the quantum anharmonic oscillator, at zero
temperature, considered in Ref.~\cite{oscillator}, while its partition
function at finite temperatures was considered in~\cite{dujo}.  {}For quantum
field theories, Ref.~\cite{jldamien} has proved convergence for a particular
perturbative series in an asymptotically-free, renormalizable model at zero
temperature. {}For a critical $\lambda \phi^4$ $O(N)$ theory in three
dimensions the issue of convergence was studied in~\cite{critical} employing
both PMS and FAC optimization criteria.  {}Finally, regarding the possible
solutions that can emerge from the optimization criteria (PMS or FAC), we must
use a definite approach in selecting the optimum root $\bar \eta$ from either
Eq. (\ref{pms}) or (\ref{fac}).  The problem of dealing with the many possible
solutions for $\bar{\eta}$ was treated in details in the first two papers
cited in Ref.~\cite{critical}, where the convergence of the LDE was also
studied in details. Typically, the higher the order in $\delta$, the more
solutions can appear. As shown in those references, all solutions at each
given order in $\delta$ can be classified into families.  The optimum value
for $\eta$ is chosen as follows: The trivial solutions for $\bar{\eta}$, e.g.
$\bar{\eta}=0$ and those that are not dependent on the coupling constant (and
thus cannot lead to nonperturbative results) are not considered. In addition
to these, at first order there is only one nontrivial solution (first family),
consistent with all our approximations, (like the high-temperature
approximation, used later in our calculations).  This family is then followed
in the next orders and used in all our calculations.  As proved in earlier
references with the LDE method, this is a consistent and unambiguous way for
choosing the optimum value for $\eta$.

It is important to stress that in the method of the LDE the selection and
evaluation of {}Feynman diagrams proceed in the same fashion as in ordinary
perturbation theory, including the renormalization
procedure~\cite{early,jldamien,critical}.  The results obtained are free from
infrared divergences, even at the critical point and in its neighborhood, thus
making it a particularly suitable method to study phase transition phenomena
in quantum field theories.  It is important to recall here that there are
similarities between the LDE and the SPT methods. In particular, the
implementation of the latter can be put in a form similar to the LDE by means
of a modified loop expansion~\cite{chiku_hatsuda}, named optimized
perturbation theory (OPT) in this reference.  But there are also some major
differences between these methods. {}For instance, in the LDE no assumption is
made {\it a priori} for the parameter $\eta$, while in the SPT/OPT it is
assumed that such a mass term is already of some order in the coupling
constant. The implication of this is that the order counting of loop expansion
has to be readjusted accordingly.

In the present paper we study the application of the LDE to the $\lambda
\phi^4$ theory. We will study the applicability of the PMS and FAC
optimization criteria for the symmetric and broken phases of the theory, and
compare results obtained with both methods. In addition, in the present work
we choose to optimize the free energy, instead of the self-energy like in many
other works employing the LDE, particularly Refs.~\cite{marcus-rudnei,mra}.
There are several reasons for doing so~\cite{KMP,GN}, but an important one is
that in some situations it might happen that the optimization of the
self-energy does not lead to nontrivial solutions, while optimization of the
free energy with PMS or FAC are seen to lead to nontrivial solutions already
at first order in $\delta$. The critical temperature $T_c$, the pressure $P$,
and the background dependent free energy $F$ are obtained here in an explicit
calculation up to order $\delta^2$. Calculations at this order require a
calculation of vacuum terms up to three loops. Since the thermodynamics of
this model has been extensively studied before in the literature with a number
of methods, our calculation here will be useful to benchmark the application
of the LDE and its two main optimization procedures against those previous
applications. In addition, we compare our results with those obtained with
standard perturbation theory.  Besides correctly reproducing the expected
second order phase transition pattern for the model, our results at order
$\delta^2$ are shown to be sufficient to obtain the thermodynamics of the
model, in the sense that the results at ${\cal O}(\delta^2)$ are not much
different from the ones at ${\cal O}(\delta)$. The results point towards a
quickly convergent LDE, as already indicated in previous studies with
different models under different conditions~\cite{jldamien,critical}.

This work is organized as follows. In the next section we introduce the
interpolation procedure for the model. In~Sec.~III we compute the free energy
in the symmetric and broken phases to ${\cal O}(\delta^2)$.  In~Sec.~IV we
present the results obtained from the optimization procedures. The pressure is
evaluated and contrasted order by order with the one obtained within
perturbation theory.  The critical temperature, the temperature dependent
vacuum expectation value of the scalar field and the free energy are
determined to ${\cal O}(\delta^2)$.  Our conclusions are presented in Sec.~V.

\section{The Model Lagrangian Density}

The interpolation defined by Eq. (\ref{opt}) when applied for the standard
$\lambda\phi^4$ model gives
\begin{eqnarray}
\mathcal{L}^{\delta }  &=&  \mathcal{L}_0(\eta) - \delta
\frac{\lambda }{4!}\phi^{4} + \delta \frac{\eta^{2}}{2}\phi^2
+ \mathcal{L}_{ct}^{\delta }\;,
\label{interp}
\end{eqnarray}
where
\begin{eqnarray}
\mathcal{L}_0(\eta)=\frac{1}{2}\left( \partial _{\mu }\phi \right) ^{2}
- \frac{m_0^{2}}{2}
\phi^{2} - \frac{\eta^{2}}{2}\phi^2\;,
\label{L0}
\end{eqnarray}
and $\mathcal{L}_{ct}^{\delta}$ is the part of the Lagrangian density carrying
the renormalization terms needed to render the model finite.  Details about
this renormalization procedure in the LDE and the explicit form for
$\mathcal{L}_{ct}^{\delta}$ are given e.g. in Ref.~\cite{marcus-rudnei} for
the case of background field dependent contributions (broken symmetry phase),
while the field independent contributions (symmetric phase) were given in
Ref.~\cite{andersen-braaten-strickland} within the context of the SPT, so we
will not repeat those same renormalization details here. One should also note
that the only ``new" terms introduced by the $\delta$-expansion interpolation
are quadratic terms and so the renormalizability of the original theory is not
changed. This means that the renormalization of the theory can be carried out
in an analogous way as in ordinary perturbation theory~\cite{marcus-rudnei}.
Specifically, the interpolation procedure given by Eq.~(\ref{interp})
introduces a new (quadratic) interaction term, with Feynman rule $i\delta
\eta^2$.  In addition to this modification, the original bare propagator,
\begin{eqnarray}
S\left( k\right) = i\left( k^{2}-m_0^{2}+i\varepsilon \right) ^{-1}\;,
\end{eqnarray}
now becomes
\begin{eqnarray}
S_\delta\left( k\right)  &=&i\left( k^{2} - m_0^2 - \eta ^{2}+
i\varepsilon \right) ^{-1}\;,
\label{modified+propag}
\end{eqnarray}
while the original quartic vertex is changed from $-i \lambda$ to $ -i \delta
\lambda$.

In the next section we will show the results for the finite temperature free
energy density using the interpolated model with the LDE at
$\mathcal{O}(\delta^{2})$. We will consider the cases of $m_0^2 = |m_0|^2$ and
$m_0^2 = - |m_0|^2$ in Eq.~(\ref{L0}), corresponding to the symmetric and
broken phases, respectively.

\section{The Finite Temperature Free Energy in the LDE to ${\cal O}(\delta^2)$}

We perform the standard derivation of the free energy~\cite{dolan_jackiw} up
to $\mathcal{O}(\delta^{2})$. With the constant field introduced through the
usual shift of the scalar field, $\phi \rightarrow \phi + \varphi$, the
Lagrangian density is rewritten as
\begin{eqnarray}
\mathcal{L}[\phi(x),\varphi]=\mathcal{L}_2[\phi(x),\varphi] +
\mathcal{L}_I[\phi(x),\varphi]\;,
\end{eqnarray}
\noindent
where $\mathcal{L}_{2}$ is the part of the Lagrangian quadratic in the fields,
\begin{eqnarray}
\mathcal{L}_{2}\left[ \phi\left( x\right) ,\varphi \right] =\frac{1}{2}
\left( \partial _{\mu }\phi\right) ^{2}-\frac{1}{2}\Omega ^{2}\phi^{2}\;,
\label{L2}
\end{eqnarray}
while $\mathcal{L}_{I}$ is
\begin{eqnarray}
\mathcal{L}_{I}\left[ \phi\left( x\right) ,\varphi \right] =-\frac{
\delta \lambda }{6}\varphi \phi^{3}-\frac{\delta \lambda }{4!}\phi^{4}\;,
\end{eqnarray}
\noindent
where in Eq.~(\ref{L2}) $\Omega^2$ is given by
\begin{eqnarray}
\Omega ^{2}=\pm m_{0}^{2}+\frac{\delta \lambda }{2}\varphi ^{2}+\left( 1-\delta
\right) \eta ^{2}\;.
\label{Omega}
\end{eqnarray}
Note that in all loop contributions the propagators will carry a mass term as
given by Eq.~(\ref{Omega}).  These terms are then expanded in $\delta$ to the
desired order, thus generating the insertions of $\eta^2$ that appear as a
consequence of the quadratic vertex introduced in Eq.~(\ref{interp}).

The free energy is
\begin{eqnarray}
{}F[\varphi]&=&F_0(\varphi) + F_{\rm 1-loop}
(\varphi)\nonumber\\
&+& \frac{1}{{\cal V}} i \ln \bigg{\langle}\exp\left\{ i\int d^4x\,
\mathcal{L}_I[\phi(x),\varphi]\right\}
\bigg{\rangle}\;,
\label{free energy}
\end{eqnarray}
where $F_0(\varphi)$ is the tree-level classical potential and $F_{{\rm
    1-loop}}(\varphi)$ is the one-loop contribution to the free energy (${\cal
  V}$ is the space volume) given by
\begin{eqnarray}
{}F_{{\rm 1-loop}}(\varphi)=\frac{1}{{\cal V}}\,i\, \ln \int d\phi\;
e^{ i\int d^4x\mathcal{L}_2 [\varphi,\phi(x)] }\;.
\end{eqnarray}
Higher loops are given by the last term in Eq.~(\ref{free energy}), with the
average $\langle\cdots\rangle$ meaning
\begin{equation}
\left\langle \cdots \right\rangle =\frac{\int \mathcal{D}\phi \;
\left( \cdots \right)\; e^{ i\int d^{4}x
\,\mathcal{L}_{2}\left[ \phi\left( x\right) ,\varphi \right]
 }  }{\int \mathcal{D}\phi \;
e^{ i\int
d^{4}x\,\mathcal{L}_{2}\left[ \phi\left( x\right) ,\varphi \right]
 }}\;.
\end{equation}
As said above, the scalar field propagators in the diagrams are obtained from
$\mathcal{L}_2[\phi(x),\varphi]$, and the vertices are determined from
$\mathcal{L}_I[\phi(x),\varphi]$, with both as given at the end of Sec. II.

Our calculations are performed, as usual, in the imaginary-time
formalism~\cite{dolan_jackiw}. Thus, the scalar boson field has Euclidean
four-momentum $P=(\omega_n,{\bf{p}})$, with $P^2=\omega_n^2+{\bf{p}}^2$, where
$\omega_n$ are the discrete Matsubara bosonic frequencies $\omega_n = 2\pi
n/\beta$, with $n=0,\pm 1, \pm 2, \cdots$, and $\beta=1/T$. Loop diagrams
involve sums over the Matsubara frequencies and integrals over the space
momentum ${\bf{p}}$. All space momentum integrals are performed in arbitrary
dimension $d=3-2 \epsilon$ and renormalization is performed in the modified
minimal subtraction scheme ($\overline{\rm MS}$).  The measure used in the
sum-integrals is then defined as

\begin{eqnarray}
\hbox{$\sum$}\!\!\!\!\!\!\!\int =
\left( \frac{e^{\gamma }\mu ^{2}}{4\pi }\right)^{\epsilon }
\beta^{-1} \sum_{n}\int \frac{d^{3-2\epsilon }\,p}{\left( 2\pi
\right)^{3-2\epsilon }}\;,
\end{eqnarray}
where $\mu$ is an arbitrary momentum scale in dimensional regularization. The
factor $\left( \frac{e^{\gamma }\mu^{2}}{4\pi }\right)^{\epsilon }$ is
introduced so that, after minimal subtraction of the poles in $\epsilon$ due
to ultraviolet divergences, $\mu$ coincides with the renormalization scale in
the $\overline{\rm MS}$ scheme.

\begin{figure}[ht]
  \includegraphics[scale=1.0]{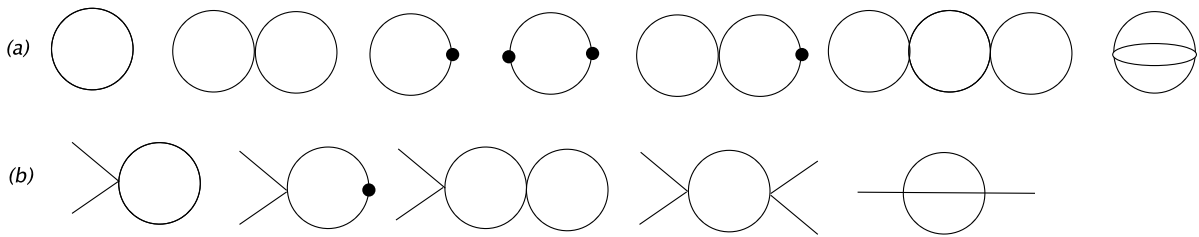}
\caption{Diagrams contributing to the free energy up to 
  $\mathcal{O}(\delta^2)$, given by (a) vacuum diagrams and (b) background
  field (external legs).  The black dots indicate a $\delta \eta^2$
  insertion.}
\label{loopdiagrams}
\end{figure}

{}From Eq.~(\ref{free energy}), the free energy is expressed up to
$\mathcal{O} \left(\delta^2\right)$ by expanding all appropriate terms in
$\delta$. Considering the vacuum contributions to the free energy, this means
that terms up to three-loops must be included. All bare (unrenormalized)
contributions are shown in {}Fig.~\ref{loopdiagrams}.

\begin{figure}[ht]
  \includegraphics[scale=0.70]{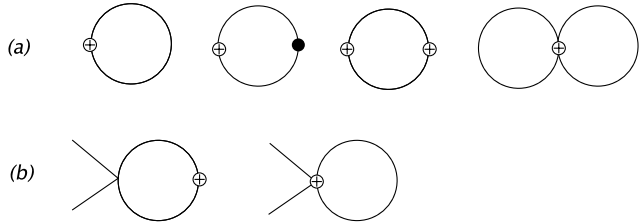}
\caption{Diagrams representing the counterterms for the free energy up to
  $\mathcal{O}(\lambda^2)$: (a) vacuum contribution, (b) background field
  contribution. As in {}Fig. 1, the black dot indicates a $\delta \eta^2$
  insertion. The circle-cross denotes either insertion of a mass counterterm
  or of a vertex counterterm.}
\label{CTdiagrams}
\end{figure}

The renormalization procedure for the symmetric phase was performed in detail
in Refs.~\cite{marcus-rudnei,andersen-braaten-strickland}.  The counterterms
for the vacuum diagrams are given in Ref.~\cite{andersen-braaten-strickland},
while those for the field dependent diagrams are given in Ref.~\cite{marcus-rudnei}. We
also note that the divergences in the broken phase can be removed by the same
counterterms determined for the symmetric phase~\cite{chiku_hatsuda,Lee,Kugo},
so the renormalization for the broken phase does not require extra effort.
The renormalization proceeds just as in standard perturbation theory and as
shown in detail in Ref.~\cite{marcus-rudnei}, only temperature independent
counterterms are required and the temperature dependent divergent terms cancel
out exactly. All diagrams of counterterms contributing to $F[\varphi]$ up to
$\mathcal{O}(\delta^2)$ are shown in {}Fig.~\ref{CTdiagrams}.

The circle-cross in {}Fig.~\ref{CTdiagrams} denotes either a mass counterterm
vertex $\Delta m^2$, or a vertex counterterm $\Delta \lambda$, given
respectively by~\cite{marcus-rudnei}
\begin{eqnarray}
&& \Delta m^2 = \delta \frac{\lambda}{32 \pi^2 \epsilon}\left[
\left(m^2 + (1-\delta) \eta^2\right) \right]
- \delta^2 \frac{\lambda^2}{(32 \pi^2)^2}\left(\frac{-2}{\epsilon^2}+
\frac{1}{\epsilon}\right)\left(m^2 + \eta^2\right)\;,
\label{Deltam} \\
&& \Delta \lambda = -\delta^2 \frac{3\lambda^2}{32 \pi^2 \epsilon}\;.
\label{Deltalambda}
\end{eqnarray}
The final expression for the renormalized free energy $F[\varphi]$, including
all terms shown in {}Figs.~\ref{loopdiagrams} and \ref{CTdiagrams} becomes
\begin{eqnarray}
F[\varphi] = F_{\rm{vacuum}} + F_{\varphi}
\label{freeenergy}
\end{eqnarray}
where $F_{\rm{vacuum}}$ denotes the vacuum contributions,
\begin{eqnarray}
F_{\mathrm{vacuum}} &=&-\frac{1}{8\left( 4\pi \right) ^{2}}\left[ 2\ln
\left( \frac{\mu ^{2}}{\mathcal{M} ^{2}}\right) +3\right] \mathcal{M}^{4}
-\frac{1}{
2\left( 4\pi ^{2}\right) }J_{0}\left( \beta \mathcal{M} \right) T^{4}
\nonumber\\
&+&\delta \frac{\lambda }{8\left( 4\pi \right) ^{4}}\left[ \left( \ln \left(
\frac{\mu ^{2}}{\mathcal{M}^{2}}\right) +1\right) \mathcal{M}^{2}-J_{1}
\left( \beta
\mathcal{M} \right) T^{2}\right] ^{2}  \nonumber \\
&+&\delta \frac{\eta ^{2}}{2\left( 4\pi \right) ^{2}}\left[ \left(
\ln \left( \frac{\mu ^{2}}{\mathcal{M} ^{2}}\right) +1\right) \mathcal{M}
^{2}-J_{1}\left( \beta \mathcal{M} \right) T^{2}\right]   \nonumber \\
&-&\delta ^{2}\frac{\eta ^{4}}{4\left( 4\pi \right) ^{2}}\left[
\ln \left( \frac{\mu ^{2}}{\mathcal{M} ^{2}}\right) +J_{2}\left( \beta
\mathcal{M}
\right) \right]   \nonumber \\
&-&\delta ^{2}\frac{\lambda }{4\left( 4\pi \right) ^{4}}\eta ^{2}
\left[ \ln\left( \frac{\mu ^{2}}{\mathcal{M} ^{2}}\right) +J_{2}
\left( \beta \mathcal{M} \right)
\right] \left[ \left( \ln \left( \frac{\mu ^{2}}{\mathcal{M} ^{2}}\right)
+1\right) \mathcal{M}^{2}-J_{1}\left( \beta \mathcal{M} \right) T^{2}\right]
\nonumber \\
&-&\delta ^{2}\frac{\lambda ^{2}}{48\left( 4\pi \right) ^{6}}\left\{
\left[ 5\ln ^{3}\left( \frac{\mathcal{M}^{2}}{\mu ^{2}}\right)
+17\ln ^{2}\left( \frac{\mathcal{M} ^{2}}{\mu ^{2}}\right) +
\frac{41}{2}\ln \left( \frac{\mathcal{M} ^{2}}{\mu^{2}}\right)
-23-\frac{23}{12\pi ^{2}}\right. \right.   \nonumber \\
&-&\left. \left. \psi ^{\prime \prime }\left( 1\right) +C_{0}
+3\left( \ln \left( \frac{\mathcal{M} ^{2}}{\mu ^{2}}\right)
+1\right)^{2}J_{2}\left( \beta\mathcal{M} \right) \right] \mathcal{M}^{4}
-\left[ 12\ln ^{2}\left( \frac{\mathcal{M} ^{2}}{\mu ^{2}}\right)
+28\ln \left( \frac{\Omega ^{2}}{\mu ^{2}}\right) \right.\right.
\nonumber \\
&-&\left. \left. 12-\pi ^{2}-4C_{1}+6\left( \ln \left(
\frac{\mathcal{M}^{2}}{\mu^{2}}\right) +1\right) J_{2}
\left( \beta \mathcal{M} \right) \right] J_{1}
\left( \beta \mathcal{M} \right) \Omega^{2}T^{2}\right.
\nonumber \\
&+&\left. \left[ 3\left( 3\ln \left( \frac{\mathcal{M}^{2}}{\mu^{2}}\right)
+4\right) J_{1}^{2}\left( \beta \mathcal{M} \right) +3J_{1}^{2}\left( \beta
\mathcal{M} \right) J_{2}\left( \beta \mathcal{M} \right) +6K_{2}
+4K_{3}\right] T^{4}\right\} \;,
\label{Fphi0}
\end{eqnarray}

\noindent
and $F_{\varphi}$ denotes the background field dependent contributions,

\begin{eqnarray}
F_{\mathrm{\varphi }} &=&F_{0}+\left\{ \frac{\delta \lambda \mathcal{M}^{2}}{
32\pi ^{2}}\left[ \log \left( \frac{\mathcal{M}^{2}}{\mu ^{2}}\right) -1+
\frac{T^{2}}{\mathcal{M}^{2}}J_{1}\left( \beta \mathcal{M}\right) \right]
-\delta ^{2}\frac{\lambda \eta ^{2}}{32\pi ^{2}}\left[ \ln \left( \frac{
\mathcal{M}^{2}}{\mu ^{2}}\right) -J_{2}\left( \beta \mathcal{M}\right)
\right] \right.
\nonumber \\
&-&\delta ^{2}\frac{\lambda ^{2}\mathcal{M}^{2}}{2\left( 32\pi ^{2}\right)
^{2}}\left[ \left( \ln \left( \frac{\mathcal{M}^{2}}{\mu ^{2}}\right)
\right) ^{2}+\frac{\pi ^{2}}{6}\right] -\delta ^{2}\frac{3\lambda ^{2}
\mathcal{M}^{2}}{2\left( 32\pi ^{2}\right) ^{2}}\left[ \left( \ln \left(
\frac{\mathcal{M}^{2}}{\mu ^{2}}\right) -1\right) ^{2}+1+\frac{\pi ^{2}}{6}
\right]
\nonumber \\
&+&\delta ^{2}\frac{\lambda ^{2}}{1024\pi ^{4}}\left[ \mathcal{M}^{2}\left(
1+\frac{\pi ^{2}}{6}\right) +4\mathcal{M}^{2}\ln \left( \frac{\mu }{\mathcal{
M}}\right) \left[ 1+J_{2}\left( \beta \mathcal{M}\right) \right]
+J_{2}\left( \beta \mathcal{M}\right) \mathcal{M}^{2}\right.
\nonumber \\
&+&\left. 8\mathcal{M}^{2}\ln ^{2}\left( \frac{\mu }{\mathcal{M}}\right)
-4\ln \left( \frac{\mu }{\mathcal{M}}\right) J_{1}\left( \beta \mathcal{M}
\right) T^{2}-J_{2}\left( \beta \mathcal{M}\right) J_{1}\left( \beta
\mathcal{M}\right) T^{2}\right]
\nonumber \\
&+&\delta ^{2}\frac{\lambda ^{2}T^{2}}{24\left( 4\pi \right) ^{2}}\left[ \ln
\left( \frac{\mathcal{M}^{2}}{T^{2}}\right) +5.3025\right] +\delta ^{2}\frac{
\lambda ^{2}\mathcal{M}^{2}}{256\pi ^{4}}\left[ \frac{\pi ^{2}}{24}-3\ln
\left( \frac{\mathcal{M}}{\mu }\right) \right.   \nonumber \\
&+&\left. \left. 2\ln ^{2}\left( \frac{\mathcal{M}}{\mu }\right) +1.164032
\right] \right\} \frac{\varphi ^{2}}{2}-\delta ^{2}\frac{3\lambda ^{2}}{
32\pi ^{2}}\left\{ \log \left( \frac{\mathcal{M}^{2}}{\mu ^{2}}\right)
-J_{2}\left( \beta \mathcal{M}\right) \right\} \frac{\varphi ^{4}}{4!}\;.
\label{Fphineq0}
\end{eqnarray}

\noindent
with $F_0$ given by
\begin{eqnarray}
{}F_0 = \frac{1}{2}\left[\pm m_0^2+(1-\delta) \eta^2 \right]
\varphi^2 + \delta \frac{\lambda}{24} \varphi^4\;.
\label{F0}
\end{eqnarray}
In Eqs.~(\ref{Fphi0}) and (\ref{Fphineq0}), $\mathcal{M}^2 = \pm m_0^2 +
\eta^2$, and the constant terms appearing in Eq.~(\ref{Fphi0}) are defined as
follows: $\psi ^{\prime \prime }\left( 1\right) =-2\zeta \left( 3\right)$,
where $\zeta(x) $ is the zeta function, $C_{2} \simeq 39.429$ and $C_{3}
\simeq -9.8424$, while $K_{2}$ and $K_{3}$ are three-dimensional integrals
that can be evaluated numerically~\cite{prd62-strickland}.  In the 
high-temperature limit, $\mathcal{M}/T \ll 1$, they are given
by~\cite{andersen-braaten-strickland}
\begin{eqnarray}
K_{2} \simeq\frac{32\pi ^{4}}{9}\left[ \ln \left( \beta\mathcal{M}\right)
-0.04597\right] - 372.65 \beta\mathcal{M}\left[ \ln \left(\beta\mathcal{M}
\right)+
1.4658\right]\;,
\end{eqnarray}
and
\begin{equation}
K_{3} \simeq 453.51+1600 \beta\mathcal{M}
\left[ \ln \left(\beta\mathcal{M} \right) +
1.3045\right]\;.
\end{equation}
In Eqs.~(\ref{Fphi0}) and (\ref{Fphineq0}), we have also defined the
temperature dependent integrals $J_n$ ($n=0,1,2$) as follows,
\begin{equation}
J_n(a ) = \frac{4 \Gamma\left(\frac{1}{2}\right)}
{\Gamma \left( \frac{5}{2} -n\right) }
\int_0^{\infty} dx \frac{x^{4-2n}}{\sqrt{x^2+ a^2}}
\frac{1}{e^{\sqrt{x^2+a^2}}-1}\;,
\label{Jn}
\end{equation}
which can be expressed as a series expansion as
follows~\cite{dolan_jackiw,lebellac,gardim,gardim+steffens}
\begin{eqnarray}
J_{0}(a) &=&
\frac{8\pi }{3}a^{3}+a^{4}\left( \ln \left( \frac{a }{4\pi }\right) +\gamma -
\frac{3}{4}\right)   \nonumber \\
&+&128\sum_{n=1}^\infty\frac{\left(-1\right)^{n}\left( 2n-1\right) !!\zeta
\left( 2n+1\right) a^{\left( 2n+4\right) }}{
32\left( n+2\right) !2^{n+1}\left( 2\pi \right)^{2n}} \nonumber \\
&-&\frac{4\pi ^{2}}{3}a^{2}+\frac{16}{45}\pi^{4}
\;,
\label{J0}
\end{eqnarray}
\begin{eqnarray}
J_{1}(a) &=&
-4\pi a -2a^{2}\left[ \ln
\left( \frac{a }{4\pi }\right) +\gamma -\frac{1}{2}\right] +\frac{
4\pi ^{2}}{3}  \nonumber \\
&-&16\sum_{n=1}^\infty\left( \frac{\left( -1\right)^{n}\left( 2n-1\right)
!!\zeta \left( 2n+1\right) a^{\left( 2n+2\right) }}
{4n!2^{n+1}\left( n+1\right) \left( 2\pi \right)^{2n}}
\right)\;,
\label{J1}
\end{eqnarray}
\noindent
and
\begin{eqnarray}
J_{2}(a) &=&
\frac{2\pi }{a}+2\ln \left( \frac{a}{4\pi }
\right) +2\gamma   \nonumber \\
&+&4\left[ \sum_{n=1}^\infty\frac{\left(-1\right)^{n}\left( 2n-1\right) !!
\zeta\left( 2n+1\right) a^{2n}}{n!2^{n+1}\left( 2\pi
\right)^{2n}}\right] \;.
\label{J2}
\end{eqnarray}
Equations~(\ref{J0})-(\ref{J2}) are all convergent in the high-temperature
limit as can be easily checked by considering a few terms in the sums in these
equations.

We should note that when optimizing the free energy, since $J_0$, $J_1$ and
$J_2$ are dependent on $\eta$, it is important to check the stability of the
results when truncating the sums in Eqs.~(\ref{J0})-(\ref{J2}). This is
particularly critical for parameter values such that $\mathcal{M}/T$ is not
much smaller than 1, a situation that requires a fairly large number of
terms in the sums.  In all results shown in the next section we have used
enough terms in Eqs.~(\ref{J0})-(\ref{J2}) so to obtain stable results for all
parameter and temperature values used.

\section{Optimization and Numerical Results}

We now turn to the application of the optimization procedures in the LDE and
show the results obtained by implementing the PMS, Eq.~(\ref{pms}), and FAC,
Eq.~(\ref{fac}).  As we explained in the introduction, the optimization
criteria are applied directly to the free energy.  The results obtained with
each optimization criterion are contrasted with each other and with those
available from other methods. This will then allow us to gauge the performance
of each optimization procedure regarding both reliability and convergence.

\subsection{Symmetric Phase}

\begin{figure}[htb]
  \includegraphics[scale=0.30]{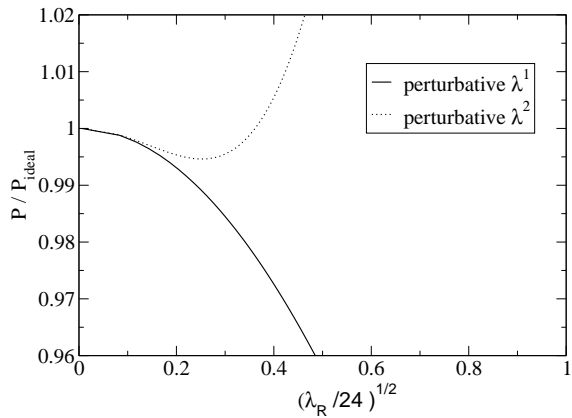}
\caption{The pressure in the symmetric phase using perturbation
  theory in $\lambda$ up to $\mathcal{O}(\lambda^2)$. The parameters used are
  $T=20\,T_0$ for the temperature and $\mu=T_0$ for the renormalization
  scale.}
\label{fig:P_pert}
\end{figure}

We initially restrict our calculations to the symmetric phase (with positive
mass term in the classical potential) and evaluate the pressure, $P = - F$.
In {}Fig.~\ref{fig:P_pert} we show our results for the pressure using the
usual perturbation theory in $\lambda$ up to $\mathcal{O}(\lambda^2)$ and
where we have restricted to the case of high temperatures ($\mathcal{M}/T \ll
1$). In this figure the behavior of the pressure is shown as a function of the
renormalized coupling constant, $\lambda_R$, and $T_0$ is a reference energy
scale chosen as $T_0=m_R(T_0)$, where $m_R$ is the renormalized mass.  This is
similar as done in Ref.~\cite{berges+reinosa} using the 2PI method. Note that
in Ref.~\cite{berges+reinosa} the authors define the quartic coupling
differently from us. In their case, $g^2 = \lambda/24$, and their results are
plotted as a function of the renormalized coupling $g_R$.  We choose here the
same scale as in Ref.~\cite{berges+reinosa} so to facilitate the comparison
with their results for the pressure. It is clear in {}Fig.~\ref{fig:P_pert}
the typical alternating behavior of the perturbative calculation, which
indicates its very poor convergence.

\vspace{1 cm}

\begin{figure}[htb]
  \includegraphics[scale=0.30]{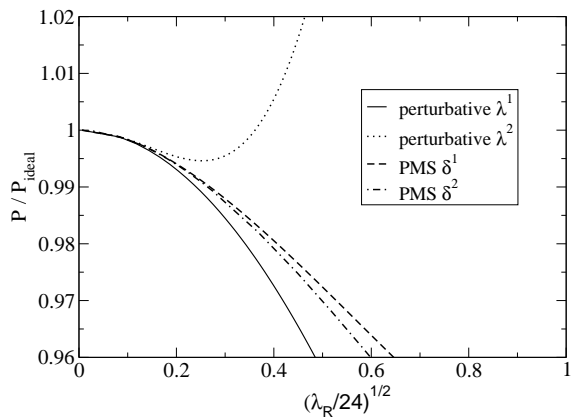}
\caption{The pressure in the symmetric phase to orders $\delta$ and
  $\delta^2$ using the PMS optimization criterion - the perturbative results
  of Fig.~\ref{fig:P_pert} are also shown. The parameters used are the same as
  those in {}Fig. \ref{fig:P_pert}.}
\label{fig:P_Pert+LDE}
\end{figure}

Next, we use the result for the free energy evaluated up to
$\mathcal{O}(\delta^2)$, given by Eq.~(\ref{Fphi0}). Note that in the
symmetric phase the pressure depends only on vacuum terms, since the free
energy is evaluated at the vacuum expectation value for the field,
$\varphi=0$. By optimizing the free energy using the PMS criterion,
Eq.~(\ref{pms}), we determine the root $\bar{\eta}$, which is then substituted
back into the expression for the free energy, with the criterion used for
choosing the optimum root as discussed below Eq.  (\ref{fac}).  This naturally
brings nonlinear $\lambda$ contributions and generates nonperturbative
results.  The pressure obtained in this case is shown in
{}Fig.~\ref{fig:P_Pert+LDE}, where we show the results obtained up to orders
$\delta$ and $\delta^2$. In the same figure we also show the perturbative
results of {}Fig.~\ref{fig:P_pert} for comparison.  It becomes evident here
the convergence of the results with the LDE-PMS, with both ${\cal O}(\delta)$
and ${\cal O}(\delta^2)$ results not differing too much, in contrast to the
perturbative (in $\lambda$) results.

\vspace{1.0 cm}

\begin{figure}[htb]
  \includegraphics[scale=0.30]{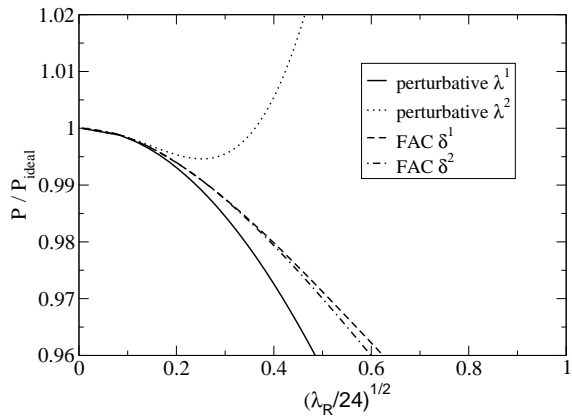}
\caption{The pressure in the symmetric phase to orders $\delta$ and $\delta^2$
  using the FAC optimization criterion.}
\label{fig:P_LDE}
\end{figure}

In {}Fig.~\ref{fig:P_LDE} we show again the results for the pressure, but now
using the FAC optimization criterion, Eq.~(\ref{fac}).  We once again see the
excellent convergence for the pressure when contrasted to the perturbative
results.

\begin{figure}[htb]
  \vspace{0.5cm} \includegraphics[scale=0.30]{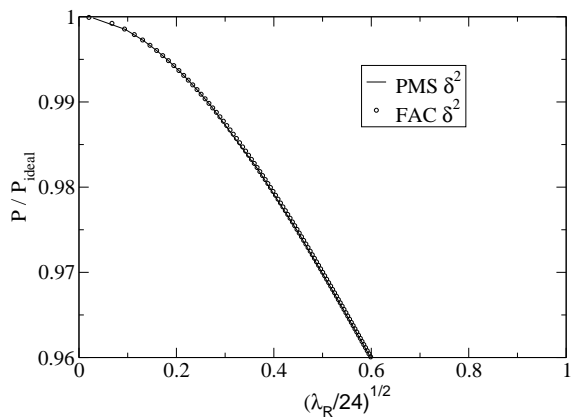}
\caption{The pressure for the symmetric phase at order $\delta^2$ using
  the PMS and FAC optimization criteria.}
\label{fig:P_FAC+PMS_d2}
\end{figure}

In {}Fig.~\ref{fig:P_FAC+PMS_d2} we plot side by side our results for the
pressure at order $\delta^2$ using the PMS and FAC optimization criteria.  It
is seen as an excellent agreement between the two optimization criteria and it
shows the equivalence of these two optimization procedures.

\begin{figure}[htb]
  \includegraphics[scale=0.30]{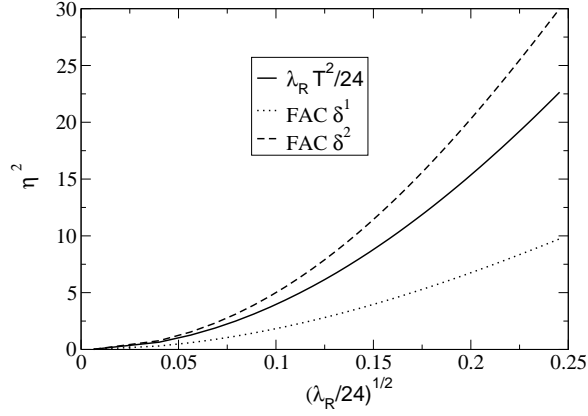}
\caption{The behavior of the optimum parameter $\eta^2$
  with respect to the renormalized coupling constant and evaluated at order
  $\delta$ and $\delta^2$ using the FAC optimization criterion.  The
  parameters used are the same as in the previous figures.}
\label{fig:eta2_LDE_d1}
\end{figure}

A side by side comparison of the order $\delta^2$ result for the pressure
(from either the PMS or FAC) with the 2PI two-loop result of
Ref.~\cite{berges+reinosa} (second panel of theirs Fig.~1) shows an excellent
agreement between the two results. Since operationally the LDE is much simpler
to be implemented than the 2PI calculation and also when compared with other
methods, like those based on the renormalization group and Schwinger-Dyson
equations, this may be a great advantage of the LDE.  Many previous
applications of the LDE to a large variety of problems (cited previously) also
confirm the strength of the method. Its strength comes basically from the fact
that its implementation is similar to that of standard perturbation theory.
The important and fundamental difference with standard perturbation theory
resides in the optimization procedure that fixes an initial, {\it a priori},
arbitrary parameter, $\eta$.  It is then interesting to investigate what kind
of role the optimum $\eta$ represents in the LDE after optimization.  This is
partially clarified in the plot shown in {}Fig.~\ref{fig:eta2_LDE_d1}, where
we show the optimum $\eta$ as a function of the renormalized coupling
constant. It shows that by increasing the order in $\delta$, $\bar \eta$
becomes closer and closer to the thermal mass $m_T$, here computed at one-loop
order for simplicity. In general, we can extrapolate this expectation and say
that the expected optimum value for $\eta$ should be the (quantum and) thermal
mass (quadratic in the field) corrections, as would be derived from a true gap
equation. This is in fact confirmed by the many applications of the LDE to the
Gross-Neveu model~\cite{GN}, in which case exact results are known (in the
large-$N$ approximation) and can then be readily compared with the results
obtained from the LDE method applied to that model.

\subsection{Broken Phase}

\begin{figure}[ht]
  \includegraphics[scale=0.30]{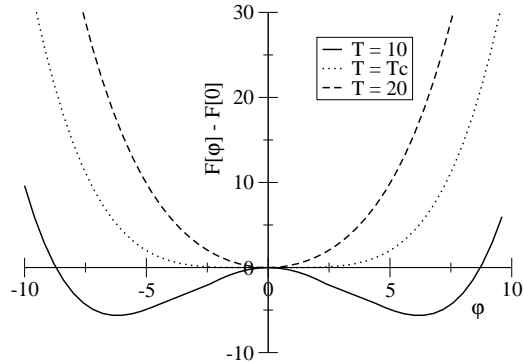} \vspace{0.5cm}
\caption{The nonperturbative free energy for $\lambda=0.1$ evaluated at
  order $\delta^2$ and using the FAC optimization criterion, for three
  different temperatures: $T<T_c$, $T=T_c$ and $T>T_c$, where $T_c=15.49$ (in
  units of the renormalization scale $\mu$). Here we have set $\mu=m_0$.}
\label{fig:veff}
\end{figure}

\begin{figure}[htb]
  \includegraphics[scale=0.30]{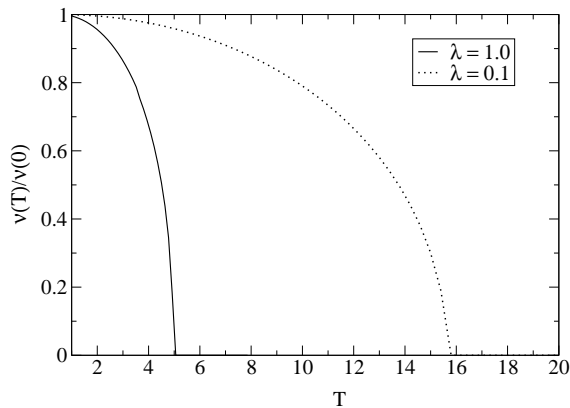}
\caption{Temperature dependence of the minimum of the free energy $\nu(T)$
  at order $\delta^2$ obtained with the FAC optimization criterion.  Here
  $\nu(0)$ is the tree-level minimum of the bare free energy.}  \vspace{0.5cm}
\label{fig:nuT}
\end{figure}

Let us now turn to the symmetry broken case (with negative square mass term in
the classical potential). {}For this case we found that only the FAC
optimization criterion leads to nontrivial solutions for $\eta$. The FAC
criterion is applied to the free energy and the resulting nonlinear equation
is solved simultaneously with the equation giving the minimum condition for
the field (thermal) expectation value, $\nu(T)$, given by the condition
\begin{eqnarray}
\left. \frac{d\, F\left[ \varphi \right] }{d\, \varphi }\right\vert
_{\varphi =\nu(T) }=0\;.
\end{eqnarray}
As it is well known, the phase transition in the pure scalar theory is second
order~\cite{arnold}, as required by universality reasons. Our results for the
free energy using the FAC criterion indicate a second order phase transition.
This is shown in {}Fig.~\ref{fig:veff}, where the free energy for $\lambda =
0.1$ is plotted for different temperature values.  The critical temperature
obtained here is $T_c/\mu \simeq 15.49$, consistent with the perturbative
prediction \cite{dolan_jackiw} and other nonperturbative calculations
\cite{Banerjee}. Another quantity that indicates that the transition is a
continuous one is the temperature evolution of the minimum of free energy,
$\nu(T)$.  This is shown in {}Fig.~\ref{fig:nuT} for $\lambda = 0.1$ and
$\lambda = 1.0$.

\vspace{1cm}

\vspace{1cm}

\begin{figure}[htb]
  \includegraphics[scale=0.30]{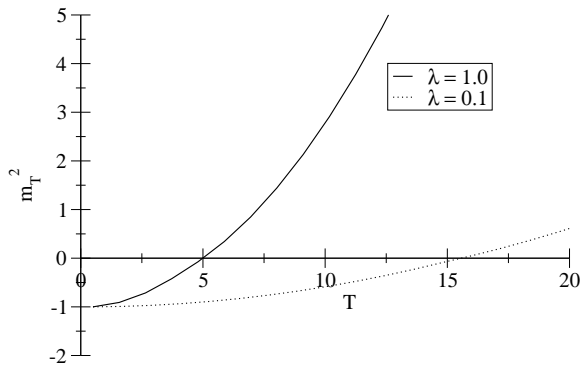} \vspace{0.5cm}
\caption{Temperature dependence of the nonperturbative thermal mass at order
  $\delta^2$ evaluated with the FAC criterion for two values of the coupling
  constant $\lambda$. We use in this plot $\mu=m_0$.}
\label{fig:mT2}
\end{figure}

{}Finally, in {}Fig.~\ref{fig:mT2} we show the temperature dependence of the
thermal mass, $m_T$, as derived from the free energy,
\begin{equation}
m_T^2 = \frac{d^2 F[\varphi]}{d^2 \varphi}\Bigr|_{\varphi=0}\;.
\label{mT2}
\end{equation}
We once again can notice a continuous and smooth transition.  We note that one
can determine the critical temperature by looking at which value of $T$
$m_T^2$ changes sign and check whether this gives the same result for $T_c$ as
obtained from $\nu(T_c)=0$ (as in {}Fig.~\ref{fig:nuT}).

\vspace{1cm}

\section{Conclusions}

One of the motivations for using the LDE to study the thermodynamics of the
scalar field theory at high temperatures, as done in this work, was its ease
of implementation and renormalization, which is no different from those of the
standard perturbation theory. One recalls that similar studies in the context
of the 2PI and related methods typically face difficulties in the
renormalization procedure, making their applicability a nontrivial task. In
addition, the LDE method, differently from other methods, like SPT (or OPT -
optimized perturbation theory), makes no assumption on the introduced mass
parameter $\eta$, thus we do not need to adjust the order of the loop
expansion accordingly.

By using the LDE, we have studied the thermodynamics of the $\lambda \phi^4$
scalar field theory in the symmetric and broken phases. The LDE is used with
two popular optimization procedures, known as PMS and FAC. There are two major
differences with the work we have done here and previous ones, like e.g. in
Refs.~\cite{marcus-rudnei,mra}.  {}First, while in general the PMS procedure
is the favorite optimization criterion in the literature related to the LDE
and similar methods, we here have shown that the FAC procedure leads to
numerically indistinguishable results from the ones obtained with the PMS. In
addition, while there may be cases where the PMS procedure leads to trivial
results only, the same may not be the case for the FAC (here we have shown
this to be the case in the broken phase). In this sense they can be used in a
complementary way, when PMS fails, one can try FAC, or vice-versa. Secondly,
unlike in Refs.~\cite{marcus-rudnei,mra}, where the quantity optimized is the
self-energy, here we choose to optimize the free energy. One advantage of this
is that, while there is no solution for the LDE at first order when optimizing
the self-energy, we do find solutions when optimizing the free energy already
at first order in $\delta$.  {}Furthermore, as shown in Ref.~\cite{GN}, the
optimization of the free energy can be shown to immediately lead to the
solution of the gap equation (here verified numerically through the results
for the optimum $\eta$), while in optimizing the self-energy further
constraints must be employed, as for example renormalization group equations.
In the numerical studies performed in the present work, we have shown that the
optimum $\eta$ carries both temperature and coupling constant contributions.
Thus, the LDE with optimization of the free energy implements automatically a
nonperturbative resummation of the thermal corrections, in conformity with
analytical results produced when this method was used to study the Gross-Neveu
model in Ref.~\cite{GN}, from which exact solutions are available and a close
comparison with the LDE results is possible.

By studying the behavior of the pressure and contrasting the results obtained
with perturbation theory and the 2PI method, we have shown that the LDE leads
to convergent results already at lowest order in the LDE expansion parameter
$\delta$, with the first and second order results changing only slightly and
producing results consistent with the 2PI nonperturbative method.  In
addition, as already mentioned above, we have shown that both optimization
procedures, FAC and PMS, lead to equivalent results for the pressure.

Another important result of our work is that the LDE is shown to be adequate
for studying phase transitions at high temperatures. In particular, when
applied to the phase transition in the $\lambda \phi^4$, the LDE predicts the
correct order of the phase transition, which is second order, in agreement
with general results of statistical mechanics. Besides this, since the LDE
method automatically produces an infrared cutoff, the results are shown to be
valid and applicable {\em below}, {\em above}, and {\em at} the critical
temperature $T_c$, showing that the LDE circumvents the usual problem seen in
perturbation theory, namely, the appearance of infrared divergences close to
critical points.

\acknowledgments

We would like to thank U. Reinosa for helpful discussions regarding their 2PI
results and the renormalization issues in the method. We would like to thank
F. Gardim for discussions on related matters.  This work was partially
supported by CNPq, FAPESP, and FAPERJ (Brazilian agencies).


\begin{thebibliography}{99}
%
\bibitem{lebellac} M. Le Bellac, {\it Thermal Field Theory} (Cambridge
  University Press, Cambridge, 1996).
%
\bibitem{GR}M. Gleiser and R. O. Ramos, Phys. Lett. {\bf B300}, 271 (1993).
%
\bibitem{revQGP}J.-P. Blaizot, E. Iancu, and A. K. Rebhan, AIP Conf. Proc.
  739, 63 (2004).
%
\bibitem{rev_dirkie}D. H. Rischke, Prog. Part. Nucl. Phys.{\bf 52}, 197
  (2004).
%
\bibitem{revrebhan}U. Kraemmer and A. K. Rebhan, Rept. Prog. Phys. {\bf 67},
  351 (2004).
%
\bibitem{revanders_strick}J. O. Andersen and M. Strickland, Ann. Phys. {\bf
    317}, 281 (2005).
%
\bibitem{Espinosa}J. R. Espinosa, M. Quir\'os and F. Zwirner, Phys. Lett.
  {\bf B291}, 115 (1992).
%
\bibitem{sato}J. Arafune, K. Ogata and J. Sato, Prog. Theor. Phys.  {\bf 99},
  119 (1998).
%
\bibitem{Camelia}G. Amelino-Camelia and S.-Y. Pi, Phys. Rev. D {\bf 47}, 2356
  (1993).
%
\bibitem{Banerjee}N. Banerjee and S. Mallik, Phys. Rev. D {\bf 43}, 3368
  (1991).
%
\bibitem{Parwani}R. R. Parwani, Phys. Rev. D {\bf 45}, 4695 (1992); erratum,
  Phys. Rev. D {\bf 48}, 5965 (1993).
%
\bibitem{zinn}J. Zinn-Justin, {\it Quantum Field Theory and Critical
    Phenomena} (Oxford University Press, 1996).
%
\bibitem{lattice}S. Muroya, A. Nakamura, C. Nonaka and T. Takaishi, Prog.
  Theor. Phys. {\bf 110}, 615 (2003).
%
\bibitem{carr} M. E. Carrington, Phys. Rev. D {\bf 45}, 2933 (1992).
%
\bibitem{arnold}P. Arnold and O. Espinosa, Phys. Rev. D {\bf 47}, 3546 (1993).
%
\bibitem{gruter+alkofer} B. Gruter, R. Alkofer, A. Maas and J. Wambach, Eur.
  Phys. J. C \textbf{42}, 109 (2005).
%
\bibitem{htl} E. Braaten and R. D. Pisarski, Nucl. Phys. \textbf{B337}, 569
  (1990); J. Frenkel and J. C. Taylor, Nucl. Phys. \textbf{B334}, 199 (1990);
  J. C. Taylor and S. M. H. Wong, Nucl. Phys. \textbf{B346}, 115 (1990).
%
\bibitem{htl12345}J. O. Andersen, E. Braaten, and M. Strickland, Phys. Rev.
  Lett. {\bf 83}, 2139 (1999); Phys. Rev. D {\bf 61}, 014017 (1999); Phys.
  Rev. D {\bf 61}, 074016 (2000); J. O. Andersen, E. Braaten, E. Petitgirard,
  and M. Strickland, Phys. Rev. D {\bf 66}, 085016 (2002); J. O. Andersen, E.
  Petitgirard, and M. Strickland, Phys. Rev. D {\bf 70}, 045001 (2004).
%
\bibitem{spt}F. Karsch, A. Patk\'os, and P. Petreczky, Phys. Lett. B {\bf
    401},69 (1997); J. O. Andersen and M. Strickland, Phys. Rev. D {\bf 64},
  105012 (2001); Phys. Rev. D \textbf{71}, 025011 (2005); J. O. Andersen and
  L. Kyllingstad, hep-ph/0805.4478.
%
\bibitem{andersen-braaten-strickland} J.O. Andersen, E. Braaten and M.
  Strickland, Phys. Rev. D \textbf{63}, 105008 (2001).
%
\bibitem{cjt}J.~M.~Cornwall, R.~Jackiw and E.~Tomboulis, Phys.\ Rev.\ D {\bf
    10} (1974) 2428.
%
\bibitem{berges+reinosa} J. Berges, Sz. Borsanyi, U. Reinosa and J. Serreau,
  Phys.  Rev. D \textbf{71}, 105004 (2005).
%
\bibitem{iancu1}J.-P. Blaizot, E. Iancu and A. Rebhan, Phys. Lett. B {\bf
    470}, 181 (1999).
%
\bibitem{iancu2}J.-P. Blaizot, E. Iancu and A. Rebhan, Phys. Rev. Lett.  {\bf
    83}, 2906 (1999).
%
\bibitem{iancu3}J.-P. Blaizot, E. Iancu and A. Rebhan, Phys.\ Rev. D {\bf 63},
  065003 (2001).
%
\bibitem{peshier} A.~Peshier, Phys. Rev. D {\bf 63}, 105004 (2001).
%
\bibitem{knoll_urko_b2_fejos}H. Van Hees and J. Knoll, Phys. Rev. D {\bf 65},
  025010 (2001); D {\bf 65}, 105005 (2002);J.-P. Blaizot, E. Iancu and U.
  Reinosa, Phys. Lett. B {\bf 568}, 160 (2003); Nucl. Phys. {\bf A736}, 149
  (2004); J. Berges, S. Borsanyi, U. Reinosa, and J. Serreau, Ann. Phys. {\bf
    320}, 344 (2005); G. Fejos, A. Patkos, and Zs. Szep, Nucl. Phys. {\bf
    A803}, 115 (2008); A. Arrizabalaga and U. Reinosa, Nucl. Phys. A
  \textbf{785}, 234 (2007); Eur. Phys. J. A \textbf{31} 754 (2007).
%
\bibitem{braaten+Peti} E. Braaten and E. Petitgirard, Phys. Rev. D
  \textbf{65}, 085039 (2002);
%
\bibitem{linear} A. Okopinska, Phys. Rev. D {\bf 35}, 1835 (1987); M. Moshe
  and A. Duncan, Phys. Lett. B {\bf 215}, 352 (1988).
%
\bibitem{GN} J.-L. Kneur, M. B. Pinto and R. O. Ramos, Phys.  Rev. D {\bf 74},
  125020 (2006); Braz. J. Phys. {\bf 37}, 258 (2007); Int.J.Mod.Phys. {\bf
    E16}, 2798 (2007); J.-L. Kneur, M. B. Pinto, R. O. Ramos and E. Staudt,
  Phys.  Rev.  D {\bf 76}, 045020 (2007); Phys. Lett. B {\bf 657}, 136 (2007);
  Int. J. Mod. Phys. {\bf E16}, 2802 (2007).
%
\bibitem{marcus-rudnei} M. B. Pinto and R. O. Ramos, Phys. Rev. D \textbf{60},
  105005 (1999).
%
\bibitem{mra} M. B. Pinto and R. O. Ramos, Phys. Rev. D {\bf 61}, 125016
  (2000).
%
\bibitem{early} V. I. Yukalov, Moscow Univ. Phys. Bull. {\bf 31}, 10 (1976);
Theor. Math. Phys. {\bf 28}, 652 (1976);
R. Seznec and J. Zinn-Justin, J. Math.  Phys. {\bf 20}, 1398
  (1979); J. C. Le Guillou and J. Zinn-Justin, Ann. Phys. {\bf 147}, 57
  (1983); V. I. Yukalov, Moscow Univ. Phys. Bull. {\bf 31}, 10 (1976); W. E.
  Caswell, Ann. Phys.  (N.Y) {\bf 123}, 153 (1979); I. G.  Halliday and P.
  Suranyi, Phys. Lett. B {\bf 85}, 421 (1979); J. Killinbeck, J. Phys.  {\bf
    A14}, 1005 (1981); R. P.  Feynman and H. Kleinert, Phys. Rev. A {\bf 34},
  5080 (1986); H. F. Jones and M. Moshe, Phys. Lett. B {\bf B234}, 492 (1990);
  A. Neveu, Nucl. Phys. (Proc.  Suppl.) {\bf B18}, 242 (1991); V. Yukalov, J.
  Math. Phys {\bf 32}, 1235 (1991); C. M. Bender, K. A. Milton and M. Moshe,
  Phys. Rev. D {\bf 45}, 1248 (1992); S. Gandhi and M. B. Pinto, Phys. Rev. D
  {\bf 46}, 2570 (1992); H.  Yamada, Z. Phys. {\bf C59}, 67 (1993); K. G.
  Klimenko, Z. Phys. {\bf C60}, 677 (1993); A.N.  Sissakian, I. L. Solovtsov
  and O. P. Solovtsova, Phys.  Lett. B {\bf 321}, 381 (1994); H. Kleinert,
  Phys. Rev. D {\bf 57}, 2264 (1998); Phys. Lett. B {\bf 434}, 74 (1998); for
  a review, see H. Kleinert and V.  Schulte-Frohlinde, {\em Critical
    Properties of $\phi^4$-Theories}, Chap.  19 (World Scientific, Singapure
  2001); M.  B. Pinto, R. O. Ramos and P. J. Sena, Physica {\bf A342}, 570
  (2004).
%
\bibitem{KMP} G. Krein, D.P. Menezes and M.B.  Pinto, Phys.  Lett. B
  \textbf{370}, 5 (1996); G. Krein, R. S. M. de Carvalho, D.P. Menezes, M.
  Nielsen and M. B. Pinto, Eur. Phys. J.  A \textbf{1}, 45 (1998); G. Krein,
  D.P. Menezes and M.B. Pinto, Braz. J. Phys. {\bf 28}, 66 (1998).
%
\bibitem{PMS}P. M. Stevenson, Phys. Rev. D {\bf 23}, 2916 (1981).
%
\bibitem{senise} M.C.B. Abdalla, J.A. Helayel-Neto, Daniel L. Nedel and Carlos
  R. Senise, Jr, Phys. Rev. D {\bf 77}, 125020 (2008).
%
\bibitem{oscillator}I.R.C. Buckley, A. Duncan and H.F. Jones, Phys. Rev. D
  {\bf 47}, 2554 (1993); C. M. Bender, A. Duncan and H.F. Jones, Phys.  Rev. D
  {\bf 49}, 4219 (1994); C. Arvanitis, H. F. Jones and C.S. Parker, Phys.
  Rev. D {\bf 52}, 3704 (1995); H. Kleinert and W. Janke, Phys. Lett. {\bf
    A206}, 283 (1995); R. Guida, K. Konishi and H. Suzuki, Ann. Phys. (N.Y.)
  {\bf 241} (1995) 152; {\it ibid.} {\bf 249}, 109 (1996); B. Bellet, P.
  Garcia and A. Neveu, Int. J. of Mod. Phys. {\bf A11}, 5587 (1996); {\it
    ibid.} {\bf A11}, 5607 (1996).
%
\bibitem{dujo}A. Duncan and H. F. Jones, Phys. Rev. D {\bf 47}, 2560 (1993).
%
\bibitem{jldamien} J.-L. Kneur and D. Reynaud, Eur. Phys. J.  {\bf C24}, 323
  (2002); Phys. Rev. D {\bf 66}, 085020 (2002).
%
\bibitem{critical}J.-L. Kneur, M. B. Pinto and R. O. Ramos, Phys. Rev. Lett.
  {\bf 89}, 210403 (2002); Phys. Rev. A {\bf 68}, 043615 (2003); E. Braaten
  and E. Radescu, Phys. Rev. Lett. {\bf 89}, 271602 (2002); Phys. Rev. A {\bf
    66}, 063601 (2002).
%
\bibitem{chiku_hatsuda} S. Chiku and T. Hatsuda, Phys. Rev. D \textbf{58},
  076001 (1998).
%
\bibitem{dolan_jackiw} L. Dolan and R. Jackiw, Phys. Rev. D \textbf{9}, 3320
  (1974); S. Weinberg, ibid. \textbf{9}, 3357 (1974); A. Linde, Rep. Prog.
  Phys. \textbf{42}, 389 (1979).
%
\bibitem{Lee} B.W. Lee, Nucl. Phys. {\textbf{B9}}, 649 (1969); B.W. Lee,
  Chiral Dynamics,(Gordon and Breach, New York, 1972)
%
\bibitem{Kugo} T. Kugo, Prog. Theor. Phys. {\textbf{57}}, 593 (1977).
  
\bibitem{prd62-strickland} J.O. Andersen, E. Braaten, and M. Strickland, Phys.
  Rev. D \textbf{62}, 045004 (2000).
%
  
\bibitem{gardim+steffens} F.G. Gardim and F.M. Steffens, Nucl. Phys. A
  {\textbf{797}}, 50 (2007).
%
\bibitem{gardim} F. G. Gardim, private communication.
%
\end{thebibliography}
\end{document}